\documentclass[namedreferences]{kluwer}
\usepackage[dvips]{graphicx}

\def \aj{Astron. J.}
\def \apj{Astrophys. J.}
\def \apjl{Astrophys. J.}
\def \apjs{Astrophys. J. Supp.}
\def \mnras{Mon. Not. Roy. Soc.}
\def \pra{Phys. Rev. A}
\def \prl{Phys. Rev. Lett.}
\def \psc{Phys. Scr.}

\newcommand{\da}{\mbox{$\Delta\alpha/\alpha$}}
\def \citet{\inlinecite}
\def \citep{\cite}

\begin{document}
\begin{article}
\begin{opening}

\title{Does the fine structure constant vary?  A third quasar
absorption sample consistent with varying $\alpha$}

\author{John K. Webb, Michael T. Murphy, Victor V. Flambaum,
Stephen J. Curran}

\institute{School of Physics, University of New South Wales,
Sydney, NSW 2052, Australia}

\begin{abstract}
We report preliminary results from a third sample of quasar absorption line
spectra from the Keck telescope which has been studied to search for any
possible variation of the fine structure constant, $\alpha$.  This third
sample, which is larger than the sum of the two previously published
samples, shows the same effect, and also gives, as do the previous two
samples, a significant result. The combined sample yields a highly
significant effect, $\da = \left(\alpha_z - \alpha_0 \right)/\alpha_0 =
-0.57 \pm 0.10 \times 10^{-5}$, averaged over the redshift range $0.2 < z <
3.7$.  We include a brief discussion of small-scale kinematic structure in
quasar absorbing clouds.  However, kinematics are unlikely to impact
significantly on the averaged non-zero \da\ above, and we have so far been
unable to identify any systematic effect which can explain it.  New
measurements of quasar spectra obtained using independent instrumentation
and telescopes are required to properly check the Keck results.

\end{abstract}


\end{opening}           

\section{Experimental status}

Any variation in $\alpha\equiv e^2/\hbar c$ would cause shifts in the
relative positions of atomic resonance transitions.  Astrophysical
measurements permit tests over cosmological time-scales through
spectroscopy of high redshift gas clouds seen in absorption against
background quasars.  Recently we introduced a new method for the
analysis of absorption systems seen in optical quasar spectra (the
{\it ``many multiplet method''}).  This technique provides an order of
magnitude increase in precision over the earlier {\it ``alkali
doublet''} method for the same quality of data
\citep{WebbJ_99a,DzubaV_99a}.

The many multiplet method compares wavelengths from many species,
exploiting the fact that the ground state levels have an enhanced
sensitivity, compared to excited levels, to any variation in $\alpha$.

The dependence of the observed wavenumber, $\omega_z$, on $\alpha$ is most
conveniently expressed as $\omega_z = \omega_0 + q \left[
\left(\frac{\alpha _z}{\alpha _0}\right)^2-1\right]$.  $\omega_0$ is the
present-day laboratory wavenumber and $\omega_z$ is the wavenumber in the
rest-frame of the absorber at the absorption redshift, $z$.  $\alpha_0$ is
the present-day value of the fine structure constant and $\alpha_z$ is the
value at the absorption redshift.

The $q$ coefficients quantify the relativistic correction for a particular
atomic mass and electron configuration.  These coefficients have been
calculated using accurate many-body theory methods
\citep{DzubaV_99a,DzubaV_99b,DzubaV_01a,DzubaV_02a}.  The parameterization
of $\omega_z$ above means that any uncertainties in the numerical values of
the $q$ coefficients will not introduce an artificial non-zero value of
\da.

\begin{figure}
\centerline{\includegraphics[width=18pc,angle=-90]{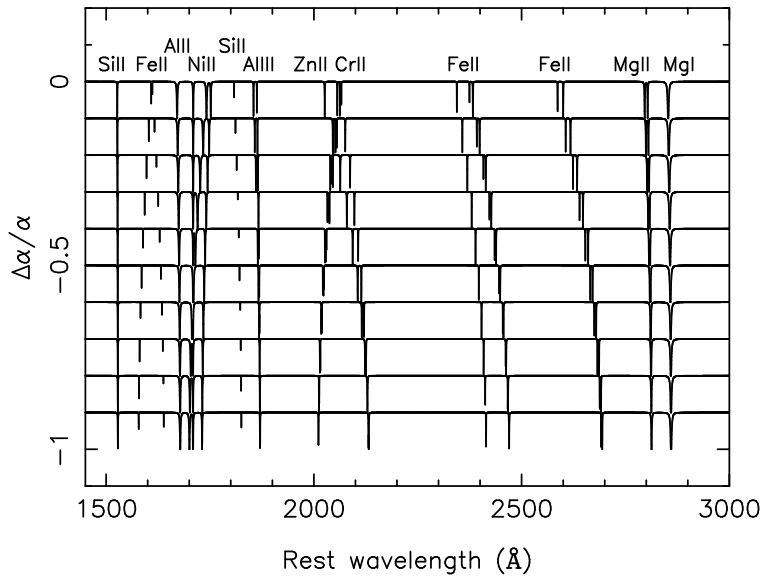}}
\caption{
Illustration of line shifts as a function of $\alpha$ for the multiplets
used in our analysis.  Note how Mg and Si act as ``anchor'' lines, and how
Zn and Cr move in opposite directions.
}\label{shifts}
\end{figure}

An important characteristic of the many-multiplet method is that a given
variation in $\alpha$ produces wavelength shifts for different species
which vary greatly in magnitude and which can be in opposite directions.
i.e. the $q$ coefficients can be of opposite sign (for the same species in
some cases) and vary by up to two orders of magnitude in numerical value
(Figure~\ref{shifts}).  This latter effect can be understood in terms of
the atomic orbital properties and the way in which $\alpha$ effects them.
This characteristic enhances the power of the many-multiplet because it
helps to reduce the impact of any systematic effects in wavelength
calibration of the data.  Since absorption clouds appear at all redshifts,
the observed transitions sample the entire observed wavelength range.  It
is therefore hard to conceive a systematic error in the wavelength
calibration which could emulate the distinctive pattern of shifts caused by
a change in $\alpha$, when a range of different transitions are included in
the fit.

In order to exploit the enhanced sensitivity defined by the $q$
values, new high precision laboratory measurements of $\omega_0$ for
the species observed were carried out using Fourier transform
spectrographs \citep{PickeringJ_98a,PickeringJ_00a,GriesmannU_00a}.
Several experiments have been completed and new highly precise
laboratory wavelengths measured for 16 transitions.  The improvement
in the accuracy of these wavelengths is $1-2$ orders of magnitude
compared to the compilation of \citet{MortonD_91a}.

Three independent samples of Keck quasar spectra have now been
analysed, covering a broad redshift range.  Two of these three samples
have been previously published.  The first sample
\citep{ChurchillC_01a} used transitions in several Mg{\sc \,ii} and
Fe{\sc \,ii} multiplets in 30 quasar absorbers.  These data provided
the first statistically significant effect in the data which was
consistent with a variation in $\alpha$ over the redshift range $0.5 <
z < 1.6$ \citep{WebbJ_99a}.

A second independent sample, of damped Lyman-$\alpha$ absorbers,
\citep{ProchaskaJ_99b} used {\it different} atomic transitions (including
Ni{\sc \,ii}, Zn{\sc \,ii}, Cr{\sc \,ii}, Al{\sc \,ii} and All{\sc \,iii})
and produced the same statistically significant non-zero result for \da\
over the redshift range $1.8 < z < 3.5$. The results of these first 2
samples are summarised in \citet{WebbJ_01a} and full analysis details are
given in \citet{MurphyM_01a}.

The analysis of the third sample (graciously provided by W. Sargent and
collaborators) is now complete and we report the preliminary results here.
Each of the three samples produces the same result and for the entire
dataset, comprising all three samples, the result is $\da \equiv (\alpha_z
- \alpha_0)/\alpha_0 = (-0.57 \pm 0.10) \times 10^{-5}$ for the redshift
range $0.2 < z < 3.7$.  We are unable to explain this result in terms of
any systematic error despite an exhaustive search \citep{MurphyM_01b}.

\section{A third independent quasar absorption line sample}

The new sample, like the previous two, was also obtained using the
HIRES spectrograph on the Keck telescope.  The spectral resolution is
${\rm FWHM} \approx 6.6{\rm kms}^{-1}$ and the signal-to-noise ratio
per pixel averages 30, with most of the sample in the range 10 to 50.
Individual exposure times ranged from 1000 to 6000\,s depending on the
quasar's apparent magnitude.  Several individual exposures were
combined to make the final spectrum.  The reader is referred to
\citet{MurphyM_01b} for specific details regarding the correction of
the wavelength scales to vacuum wavelengths.

78 absorption systems were identified which contain a sufficiently
large sample of transitions such that the many-multiplet method can
provide meaningful constraints on \da.  The redshift range covered
by this sample is $0.2 < z < 3.7$.  As with the previous two samples,
this broad redshift coverage means that \da\ is derived using different
sets of transitions at different redshifts.

\begin{figure}
\centerline{\includegraphics[width=24pc]{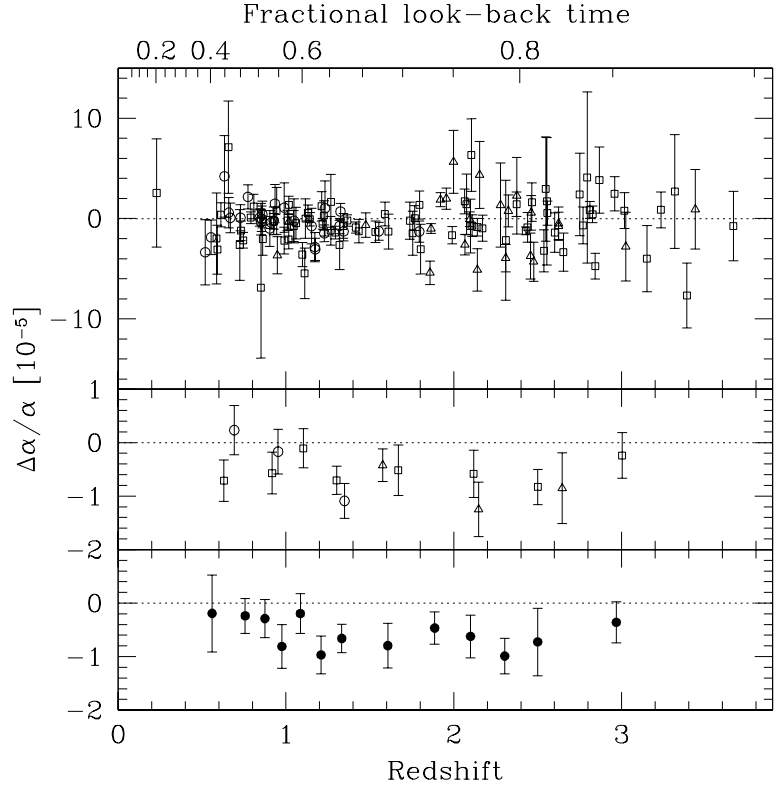}}
\caption{All results from the many-multiplet method to date.
$\Delta\alpha/\alpha$ for all 3 samples combined.  Hollow circles
(Churchill), triangles (Prochaska and Wolfe), squares (Sargent). Upper
panel: unbinned individual values.  Middle panel: binned but different
symbols still reveal results for each sample.  Lower panel: binned
over the whole sample.}
\label{cps}
\end{figure}

\section{Systematic effects and numerical analysis considerations}

Here we begin a discussion of systematic effects.  The discussion
continues in the companion paper in this volume, \citet{MurphyM_03b}.

\subsection{Wavelength distortions}

The simplest systematic problem one can imagine is a low order
distortion in the wavelength scale due to errors in the wavelength
calibration procedure.  Quasar exposures at the telescope are
bracketed in time by exposures of a standard calibration source,
ThAr.  The ThAr laboratory wavelengths are taken from
\citet{PalmerB_83a} and \citet{NorlenG_73a} and the measurement errors
in the individual ThAr wavelengths are $\sim 5\times
10^{-5}$\AA. Assuming no systematic trends, and taking the
literature-quoted errors, these errors are too small by a factor of
$\sim 40$ to produce a \da\ at the level we observe.

However, any mistakes made during the process of transferring wavelength
information from the ThAr to quasar exposures could in principle emulate a
non-zero \da.  To check this, we repeat a test previously applied to
samples 1 and 2, analysing the ThAr emission line spectra in the same way
as the quasar spectra \citep{MurphyM_01b}.  For every transition in every
quasar absorption system, we identify individual ThAr emission lines in the
corresponding calibration spectra which fall close to the observed quasar
absorption line wavelengths.  These emission lines are then fitted with
Gaussian profiles instead of Voigt profiles.  We parameterize the
``observed'' ThAr wavelength using the same relation given in Section 1 in
this paper, using the known ThAr laboratory wavelength for $\omega_0$, but
still using the $q$ coefficients from the quasar transition.  We then fit
the ThAr datasets, which directly sample the same wavelength regions as do
the quasar transitions, for all absorption systems in all quasars.  In this
way, the ThAr spectra become ``fake quasar spectra'', and provide a direct
test on the reliability of the wavelength calibration procedure.

\begin{figure}
\centerline{\includegraphics[width=24pc]{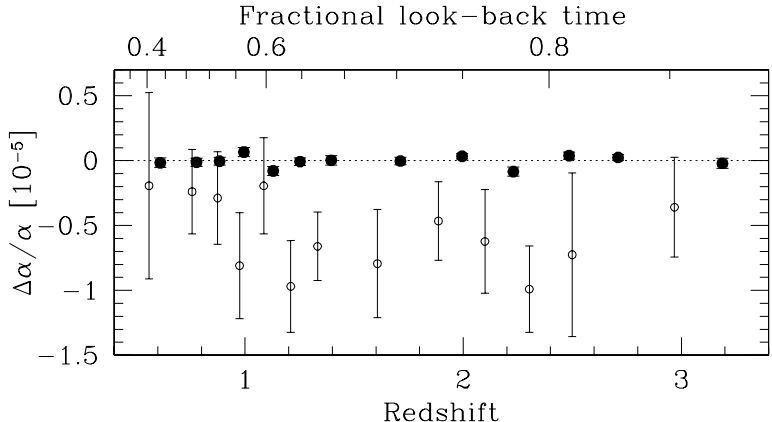}}
\caption{A check on the accuracy of the wavelength calibration.  The
hollow circles show the binned quasar results.  The filled circles
show the inferred values using the ThAr calibration spectra as ``fake
quasar spectra''.  Wavelength calibration errors do not produce the
signal seen in these quasar data.}
\label{cps_thar_short}
\end{figure}

The results are illustrated in Figure~\ref{cps_thar_short}.  The
``\da'' for the ThAr points are plotted on the same scale as the
quasar results.  This test demonstrates conclusively that the
calibration process itself introduces no significant errors.

\subsection{Minimising the number of free parameters}
\label{minimise}

Although the analysis methods have been applied extensively in other
contexts, and have also been described elsewhere in detail as applied to
\da, we re-iterate some of the features of the numerical methods which are
of particular importance here.  The absorption systems are essentially
always found to comprise multiple components, spread over 10s to 100s of
kms$^{-1}$.  Two assumptions are made in order to reduce the number of free
parameters in the fitting process, and hence to maximise the potential
precision of the measurement: {\it (i)} we adopt the same redshift (as a
free parameter) for corresponding components in different species, {\it
(ii)} we assume that the line broadening mechanisms for different species
are either entirely thermal, or entirely turbulent.  Together these amount
to the assumption of no spatial or velocity segregation for the species
used in estimating \da.

As a specific example of the reduction in the number of free
parameters, one single absorption line from species $i$ (e.g.  Mg{\sc
\,ii} 2796\AA) is defined by 3 free {\it cloud} parameters: $z_i$,
$b_i$, and $N_i$.  $z_i$ is the absorption redshift, $b_i$ is the
parameter which describes the atomic velocity dispersion and is
related to the rms distribution, $v$, by $b = \sqrt 2 v$ (kms$^{-1}$),
and $N_i$ is the column density of absorbing atoms (atoms\,cm$^{-2}$).
There are additional {\it physics} parameters required to describe the
line.  These include the rest-frame wavenumber of the species,
$\omega_z$ (determined by \da\ and $\omega_0$ as discussed in Section
1), the oscillator strength and Einstein coefficient for that
transition.

If we now include one additional line in the fit (e.g. Fe{\sc \,ii}
2382\AA) then without any assumptions we clearly double the number of
free parameters.  However, assumptions {\it (i)} and {\it (ii)} above
allow us to reduce this number from 6 to 4 by tieing the $z$'s and
relating the $b$-parameters by $b_{obs}^2 = b_{thermal}^2 + b_{bulk}^2
= \frac{2kT}{m} + b_{bulk}^2$.  In practice, we carry out 2 fits to
each absorption system, one at each of the two extreme cases, $T = 0$
(so all species are fitted with the same value of the Doppler
parameter, $b_{obs}$), and $b_{bulk} = 0$ (in which case the
$b_{obs}^2$ for each species are inversely proportional to the atomic
mass).  We require a consistent solution for \da\ for each fit,
otherwise the system is rejected.

A further important technical detail is that the numerical method
ensures that the actual constraints on \da\ are derived in a quite
natural way from {\it optically thin} species and not from saturated
absorption lines.  {\sc vpfit} is a non-linear least-squares method.
Parameters are estimated using the first and second derivatives of
$\chi^2$ with respect to the free parameters.  Where multiple
components are present, blended in a saturated absorption feature, the
value of $\chi^2$ is insensitive to relatively large changes in the
component redshifts.  Put another way, the derivatives of $\chi^2$
with respect to those redshifts are very small compared to the
optically thin case.  The consequence of this is that saturated lines
contribute little to the overall solution for \da.  This is important
to realize because the potential systematic effects due to kinematic
cloud structure and possible isotopic abundance evolution may be
larger for saturated lines.

We also note that error estimates on \da\ are derived from the diagonal
terms of the covariance matrix (inverse of the Hessian matrix) at the
best-fit solution.  This is a standard technique used in a broad range of
applications.  The implicit assumption made in ignoring off-diagonal terms
is that separate parameters are not closely correlated.  This turns out to
be a reasonable assumption for \da\ (Figure 1 illustrates why \da\ is not
degenerate with redshift).  Monte-Carlo methods have also been used to
verify this approximation.

Assumptions {\it (i)} and {\it (ii)} are addressed in the next
section.

\subsection{The effect of cloud kinematics and dynamics on \da}

For a single absorption system, a non-zero \da\ could be emulated by
velocity segregation of the different species observed.  This could arise
through chemical abundance gradients combined with differential velocity
fields.  These effects could generate departures from Voigt profiles.
However, over an ensemble of random sight-lines, the shifts in \da\ are
themselves random, and kinematic effects act just to increase the scatter
in a plot such as Fig.\ref{cps}.  We now give a brief discussion of the
contributory factors to kinematic effects on small velocity scales.

\subsubsection{Small-scale velocity structure}

The large-scale properties of the gas through which the sight-line to the
background quasar passes have no influence on estimates of \da.  Galactic
rotations or large-scale galactic winds are unimportant.  The simple
empirical fact that we get excellent agreement between the redshifts of
individual components within a quasar absorption complex, down to a small
fraction of a kms$^{-1}$ illustrates this.  Any contribution to the scatter
in \da\ comes only from the detailed properties of gas on scales smaller
than a few kms$^{-1}$.

Given the importance of small-scale properties in the determination of \da,
it is relevant to ask whether the gas is in equilibrium on these scales,
and if not, how the assumption of a Maxwellian atomic velocity distribution
inherent in fitting Voigt profiles impacts on \da\ measurements.  At the
lower redshift end of our sample, a recent detailed study (Churchill, in
preparation) suggests that the ratio of the column densities of Mg{\sc
\,ii} and Fe{\sc \,ii} appears not to change systematically across an
absorption complex.  A similar chemical uniformity is found for DLAs
\citep{ProchaskaJ_02d}. The DLAs show no significant evolution in their
number density per unit redshift interval \citep{RaoS_00a}.
Photo-ionization models applied to the data suggest gas in photo-ionized
equilibrium with an ambient extra-galactic background.  If this is correct,
local equilibrium may be valid, and any redshift evolution of the number of
absorption lines per unit redshift interval over and above that expected
due to cosmology alone for the Mg{\sc \,ii}/Fe{\sc \,ii} absorbers could be
explained by cosmological evolution in the integrated background UV flux.

We learn from all the above that there do not seem to be any gross changes
in physical conditions over large scales, across an absorption complex.
This may imply that we should not expect to find abundance variations and
non-equilibrium small-scales.

\subsubsection{Disk and halo models}

In all cases, the \da\ constraints come only from optically thin
components.  Saturated lines are effectively given zero weight in the
parameter estimation procedure as discussed in Section \ref{minimise}.
From detailed kinematic studies of quasar absorbers
\citep{BriggsF_85a,LanzettaK_92a,WolfeA_00b,ChurchillC_01a}, it seems
that a disk+halo model provides a reasonable description of the
observations.  Alternative models exist, including multiple merging
clumps bound to dark matter halos
\citep{HaehneltM_98a,McDonaldP_99a,MallerA_99a} and outflows from
supernovae winds \citep{NulsenP_98a,SchayeJ_01a}.

For disk+halo models of Mg{\sc \,ii}/Fe{\sc \,ii} systems, the disk
component is generally strongly saturated and spread over small
velocity scales.  The halo component is more broadly spread in
velocity space and causes the lower column density absorption, so the
constraints on \da\ from Mg{\sc \,ii}/Fe{\sc \,ii} most probably come
from the outer parts of galaxy halos.  The constraints on \da\ from
DLAs arise either from low abundance and hence unsaturated species
(e.g. Ni{\sc \,ii}, Cr{\sc \,ii}, Zn{\sc \,ii}), or they come from
optically thin components which flank the saturated components
(e.g. Fe{\sc \,ii}1608 or Al{\sc \,ii}).  We note in passing that any
dependence of $\alpha$ on the local gravitational potential could
conceivably emulate a cosmological evolution with redshift if DLAs
arise in or closer to galaxy disks than the Mg{\sc \,ii}/Fe{\sc \,ii}
systems.

\subsubsection{Comparison with the ISM}

Analogy with the interstellar medium in our own Galaxy does however suggest
that non-equilibrium could apply on very small scales.  In an interesting
study, \citet{AndrewsS_01a} use stars in a background globular cluster,
M92, to probe the kinematics on scales defined by the separation between
the lines of sight at the absorber.  They find significant variations in
Na{\sc \,i} column densities in the ISM on scales as small as 1600\,AU (or
$\sim 0.01{\rm \,pc}$).  Even smaller scale details of the ISM come from
measurements of temporal variation of Na{\sc \,i} and K{\sc \,i} absorption
lines, implying non-equilibrium scales $\sim 10 - 100{\rm \,AU}$
\citep{CrawfordI_00a,LauroeschJ_00a,PriceR_00a}.  These ISM sizes are small
compared to estimates of the sizes of individual cloud components,
$\sim$10--100\,pc \citep{ChurchillC_01a}.  However, it should also be noted
that gas densities appear to be quite different and so the comparison 
should be treated with caution.

\subsubsection{Quasar emission region size and cloud structure}

There may also be important geometric issues.  A characteristic size
for a quasar continuum emission region may be $\sim 10^{-3}{\rm
\,pc}$.  The intervening ISM cloudlet ``sees'' an illuminating source
which is thus not quite a perfect point source and the light rays from
the edges of the quasar continuum region pass through slightly
different regions of space at the absorber.  Consider a quasar at
$z_{em}=1$ and an absorber at $z_{abs}=0.5$.  The transverse
separation between the light rays at the absorber are only about 25\%
smaller than the emission region (this is cosmology dependent, but the
argument here is illustrative).  Thus, the applicability of Voigt
profiles, which assume a Maxwellian atomic velocity distribution,
depends on the kinematics and dynamics of the ISM in the intervening
galaxy on scales of $\sim 10^{-3}{\rm \,pc}$.

\subsubsection{Gravity}

Some theories argue that gravity may be important for cloud confinement on
small scales in the Galactic ISM \cite{WalkerM_98a}.  If gravity plays a
significant role in quasar absorption systems on similar velocity scales,
we could apply a simple stability condition, $v^2 = GM/R$ where $M$ and $R$
are the cloudlet mass and radius.  Estimates for cloudlet sizes and radii
vary but adopting $R=10{\rm \,pc}$ and $M=30\,M_{\odot}$
\citep{ChurchillC_01a} we get $v \approx 0.1{\rm \,kms}^{-1}$.  In this
case, we get an upper limit on the velocity shift between different
species.  For one single Mg{\sc \,ii}/Fe{\sc \,ii} absorber, this velocity
shift translates into an error on an individual \da\ measurement of roughly
half the detected effect.  However, this would be randomized over $\sim$100
observations, producing a maximum effect which is 20 times smaller than
that observed.

\subsubsection{Observed scatter in Fig.\ref{cps}}

The \da\ points in Fig.\ref{cps} may give some hint of kinematic
effects.  At lower redshift, the scatter in the \da\ points is
consistent with the statistical error bars, which are derived on the
basis of the signal-to-noise ratio and spectral resolution, and assume
that the best-fit solution is the correct one.  Any particular fit for
a set of multiple components to an absorbing complex is not unique.
Missing weak components will frequently or perhaps always be present.
It is extremely unlikely however that these components can
systematically affect \da\ when averaged over a large sample.  The
effect of missing components will be to increase the random scatter in
the individual \da\ values.

At higher redshift, i.e. for the DLAs, this seems to happen to a
greater extent than for the lower redshift points.  This is to be
expected for the following reasons: (a) at higher $z$ (the DLAs), a
larger number of different species (e.g. Si{\sc \,ii}, Fe{\sc \,ii},
Ni{\sc \,ii}, Zn{\sc \,ii}, Cr {\sc \,ii}) are generally available for
fitting compared to lower $z$ (i.e. Mg{\sc \,ii} and Fe{\sc \,ii}).
The range in optical depths for corresponding velocity components for
the DLAs is significantly larger than for the lower $z$ Mg{\sc
\,ii}/Fe{\sc \,ii} systems.  Abundance and ionization variations are
therefore likely to be more noticeable.  In other words, in general,
inter-comparing more species is likely to lead to greater scatter in
\da; (b) if DLAs have a more complex velocity structure, i.e. the
number of absorbing components per kms$^{-1}$ may be higher than for
the Mg{\sc \,ii}/Fe{\sc \,ii} systems, deblending would be more
difficult, increasing the uncertainty on \da; (c) with the DLAs, we
have more low-$b$ species compared to Fe{\sc \,ii}/Mg{\sc \,ii}.
These features are closer to the resolution of the instrument, so
there is an increased systematic bias against finding the weaker
components.

\section{Summary and the next step}

We have now analysed three large samples of high quality quasar
spectra.  All three produce a significant effect in the data which is
consistent with varying $\alpha$.  None of the systematic effects so
far identified explain the effect.  Kinematic effects will be
randomized over a large sample such as this, but will contribute to
the overall scatter in the individual \da.  The discussion above gives
a preliminary guide to the contributory factors, which need
quantifying in detail, in order to fully understand the scatter in the
\da\ points over and above pure statistics.

The fact that the higher redshift points exhibit more scatter suggests our
precision is close to being limited by systematics.  This in turn suggests
that further observations of quasar spectra should probably aim at
increasing the number of absorption systems studied, rather than increasing
the signal-to-noise.  However, it is important to study any possible
departures from Voigt profiles in detail, and a targeted study at the
highest possible signal-to-noise and resolution is important for this goal,
for a small number of quasars.

All three samples come from Keck/HIRES.  Whilst we have no reason to
suspect an instrument-dependent effect, this must obviously be checked.
Observations with VLT and Subaru will enable this.

Benefitting from the extra-solar planetary developments, using iodine
cells is an obvious way to completely remove any issues concerning the
different optical paths taken by quasar and calibration lamp-light,
and may eventually provide the acid-test.

\acknowledgements 

This work relies on the superb quasar spectroscopic observations of
Chris Churchill and collaborators, Art Wolfe and Jason Prochaska, and
Wal Sargent and collaborators.  The data used in this work corresponds
to a mammoth effort on their part and we very gratefully acknowledge
their crucial contributions.  We also thank John Barrow, Chris
Churchill, Joe Wolfe and Frank Briggs for useful discussions.  It is a
pleasure to thank Graca Rocha and Carlos Martins for organizing the
enjoyable conference at which this work was presented, not forgetting
Taylors Port for the hangover after the conference dinner.  We
gratefully acknowledge the John Templeton Foundation for financially
supporting this project.


\end{article}
\end{document}